\documentclass[notitlepage,twocolumn,prl,tightenlines,superscriptaddress,showpacs,floatfix]{revtex4-1} %nofootinbib %floatfix

\usepackage{amsmath,amssymb,amsfonts,latexsym}
\usepackage[colorlinks=true, linkcolor=blue, citecolor=blue]{hyperref}
\usepackage{bbm}
\usepackage[mathcal]{euscript}
\usepackage{graphicx}
\usepackage{epsfig}
\usepackage{color}
\usepackage[activate=normal]{pdfcprot}  % richtiges character protruding

\usepackage{psfrag}
\usepackage{txfonts,pxfonts,wasysym,}
\usepackage{pifont}
\usepackage[caption = false]{subfig}
\newcommand{\eccm}{Gulliver UMR CNRS 7083, ESPCI Paris, PSL Research University, 10 rue Vauquelin, 75005 Paris, France}
\newcommand{\oxford}{Physical and Theoretical Chemistry Laboratory, Department of Chemistry - University of Oxford, South Parks Road, Oxford OX1 3QZ, United Kingdom.}

\newcommand{\be}{\begin{equation}}
\newcommand{\ee}{\end{equation}}
\newcommand{\ben}{\begin{equation*}}
\newcommand{\een}{\end{equation*}}
\newcommand{\ba}{\begin{eqnarray}}
\newcommand{\ea}{\end{eqnarray}}

\newcommand{\rr}{\boldsymbol{r}}
\newcommand{\nn}{\boldsymbol{n}}
\newcommand{\vv}{\boldsymbol{v}}

\newcommand{\PPi}{\boldsymbol{\Pi}}

\begin{document}
%\graphicspath{{./figures/}}

\title{Interrupted Motility Induced Phase Separation in Aligning Active Colloids}

\author{Marjolein N. van der Linden}
\affiliation{\eccm}
\affiliation{\oxford}
\author{Lachlan C. Alexander}
\affiliation{\oxford}
\author{Dirk G.A.L. Aarts}
\affiliation{\oxford}
\author{Olivier Dauchot}
\affiliation{\eccm}

\date{\today}

\begin{abstract}
Switching on high activity in a relatively dense system of active Janus colloids, we observe fast clustering, followed by cluster aggregation towards full phase separation. The phase separation process is however interrupted when large enough clusters start breaking apart. Following the cluster size distribution as a function of time, we identify three successive dynamical regimes. Tracking both the particle positions and orientations, we characterize the structural ordering and alignment in the growing clusters and thereby unveil the mechanisms at play in these regimes. In particular we identify how alignment between the neighboring particles is responsible for the interruption of the full phase separation. This experimental study, which provides the first large scale observation of the phase separation kinetics in active colloids, combined with single particle analysis of the local mechanisms, points towards the new physics observed when both alignment and short-range repulsions are present. 
\end{abstract}

\maketitle
It has been widely reported, both experimentally~\cite{Deseigne:2010gc,Cates:2010hx,Peruani:2012dya,SchwarzLinek:2012ksa,Theurkauff:2012jo,Palacci:2013eua,Buttinoni:2013dea,Liu:2017uk,Ginot:2018ff} and numerically~\cite{Peruani:2006gl,Peruani:2010jh,Peruani:2011kwa,Fily:2012hj,Farrell:2012ks,Fielding:2012vt,Cates:2013ia,Redner:2013jo,Stenhammar:2013kb,Mognetti:2013jza,Fily:2014cp,Levis:2014ina,Stenhammar:2014fc,Weitz:2015bca,Mani:2015ku,Tung:2015hq,Patch:2017be,Alarcon:2017gc,MartinGomez:2018bwa,SeseSansa:2018wl,Shi:2018tp}, that self-propelled particles show a strong tendency to phase separate or form clusters with various structural and dynamical properties.
Two limiting scenarios have been identified. When alignment dominates the interactions, a transition to polar or nematic order takes place following a phase separation between a disordered gas and an orientationally ordered liquid. At coexistence, polar bands or nematic lanes dominate the dynamics. This physics is captured in Vicsek-like models, where constant-speed point particles align their velocities according to effective rules~\cite{vicsek1995novel,Chate:2006en,Chate:2008isb,Solon:2014tu}. When excluded volume interactions dominate and crowding effects slow down the propulsion speed, a motility-induced phase separation (MIPS) takes place: coarsening leads to the formation of one large droplet surrounded by a disordered gas phase~\cite{Fily:2012hj,Cates:2013ia,Stenhammar:2013kb}. Both scenarios are well understood at the level of large-scale hydrodynamic equations~\cite{toner1998flocks,Peshkov:2014un,Cates:2015ft}.

In experimental situations, clustering results from the interplay of several factors such as self-propulsion, excluded volume, alignment and noise, in addition to usual attractive, repulsive and hydrodynamic interactions. Disentangling these effects is a truly challenging task~\cite{Ball:2013ck}, which has motivated a large number of numerical studies~\cite{Mognetti:2013jza,Tung:2015hq,Mani:2015ku,Patch:2017be,Alarcon:2017gc,MartinGomez:2018bwa,SeseSansa:2018wl}. 
%In the absence of alignment, the situation is now pretty clear: at intermediate densities and rather low noise~\cite{Buttinoni:2013dea}, the dynamics is governed by the MIPS scenario; conversely when clusters develop at low densities and moderate noise level~\cite{Theurkauff:2012jo,Palacci:2013eua}, they are triggered by weak local attraction, eventually amplified by activity~\cite{Mani:2015ku}.
Of particular interest, is the case where alignment and excluded volume are simultaneously present. These are the minimal ingredients at play in the population dynamics of elongated micro-organisms\cite{BenJacob:2000ei,Kaiser:2003ef,Dombrowski:2004eu,Cates:2010hx,Liu:2013hj,Liu:2017uk}. On one hand, it was argued that alignment reduces the rotational diffusion and therefore favors MIPS~\cite{SeseSansa:2018wl,MartinGomez:2018bwa}. On the other hand, recent simulations of self-propelled rods suggest that steric alignment reduces MIPS to a minor part of the phase diagram~\cite{Shi:2018tp}, in agreement with earlier simulations, which had reported the existence of complex dynamical phases~\cite{Peruani:2011kwa,Weitz:2015bca}.

% -- FIGURE 0001 ------------------------------
\begin{figure}[t!]
\vspace{-0mm}
\includegraphics[width=0.95\columnwidth,trim = 0mm 0mm 0mm 0mm, clip]{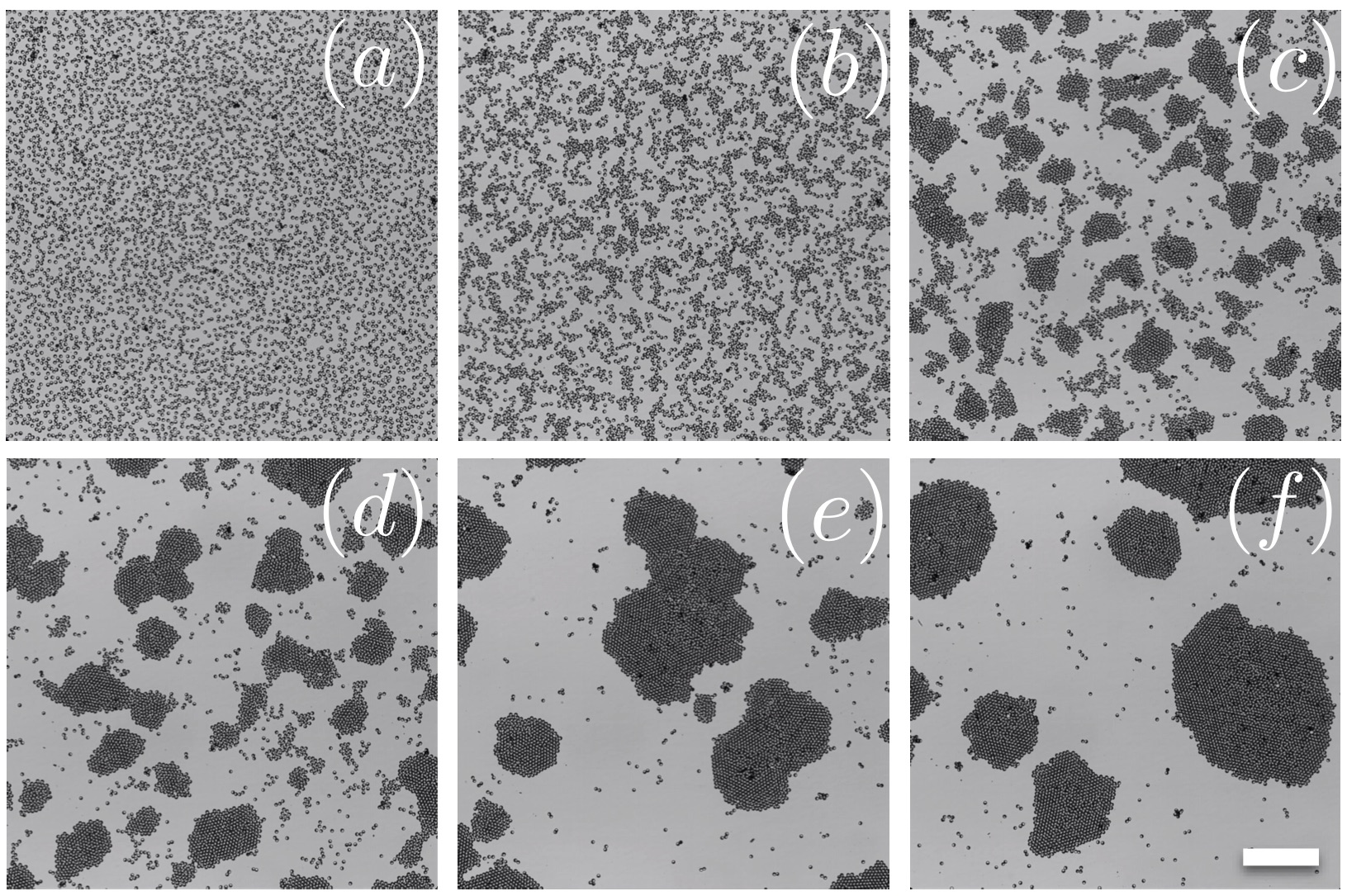}
\vspace{-0mm}
\caption{{\bf Aggregation kinetics in a system of induced-charge electrophoretic self-propelled Janus colloids:} From (a) to (f)~: successive time steps ($t=0.02; 0.4; 2; 5; 20; 68$ s) following the onset of activity. Scale bar is $100\mu$m. See also Movie-1 in Supp. Mat.}
\label{fig:coarsening}
\vspace{-5mm}
\end{figure}
%
%To our knowledge, there has been no experimental investigation of this situation of broad interest in a well controlled system of synthetic micro-swimmers. 
In this letter, we take advantage of a 2D experimental system of induced-charge electrophoretic self-propelled Janus colloids~\cite{Nishiguchi:2015dy,Yan:2016ko} to study the clustering and coarsening processes (Fig.~\ref{fig:coarsening}). We specifically focus on the aggregation kinetics and demonstrate that (i) initially the cluster size rapidly increases, with alignment playing no role; (ii) a second regime of aggregation takes place, during which cluster dynamics, composed of rigid body translation and rotation, is dominated by orientational ordering of colloids inside the clusters; (iii) the coarsening is slowed down by fragmentation events until the average cluster size eventually stops growing and a steady state is reached. The structural and polar ordering within the clusters reveals that the largest clusters break up along grain boundaries, which were formed during their aggregation. These regions populated with defects cannot resist the active stresses resulting from the alignment within the grains.   

The experimental system, following~\cite{Yan:2016ko}, is composed of tens of thousands of Janus colloids (silica particles with a diameter $d$ of $4.28$ $\mu$m half-coated in $35$ nm of titanium followed by $15$ nm of silica) in an aqueous solution of $0.1 mM$ NaCl, sandwiched between two ITO coverslips (Diamond Coatings) that were coated with $25$ nm of silica, separated by $\sim95 \mu$m thick spacers. The particles form a monolayer, with a surface fraction $\phi\simeq 0.25$, on the bottom electrode. When a square wave with a frequency of $10$ kHz and an amplitude of $10$ V is applied, the particles self-propel with their silica side facing forward, as prescribed by induced-charge electrophoresis (ICEP)~\cite{Gangwal:2008ge}. 
We record the  dynamics at $50$ fps using an Olympus Plan N 20x/0.40 objective and 2048x2048 pixels camera. This allows us to capture the large scale dynamics, while simultaneously tracking the particles positions $\rr_k(t)$ and orientations $\nn_k(t)$. The nominal velocity of an individual particle is $v_0\simeq 20d/$s. For such large particles, with negligible rotational diffusion, this leads to very persistent trajectories. At the working frequency the dielectric dipole-dipole interactions are weak~\cite{Yan:2016ko}. We however notice a short-range repulsion, together with a small head-to-tail attraction. The latter does not resist multiple collisions and does not lead to the formation of long chains as reported in~\cite{Yan:2016ko}, at much higher frequency. The total number of particles $M$ inside the field of view remains approximately constant ($M\simeq 5500$). 

\begin{figure}[t] 
\center
\vspace{0cm}
\includegraphics[width=0.9\columnwidth]{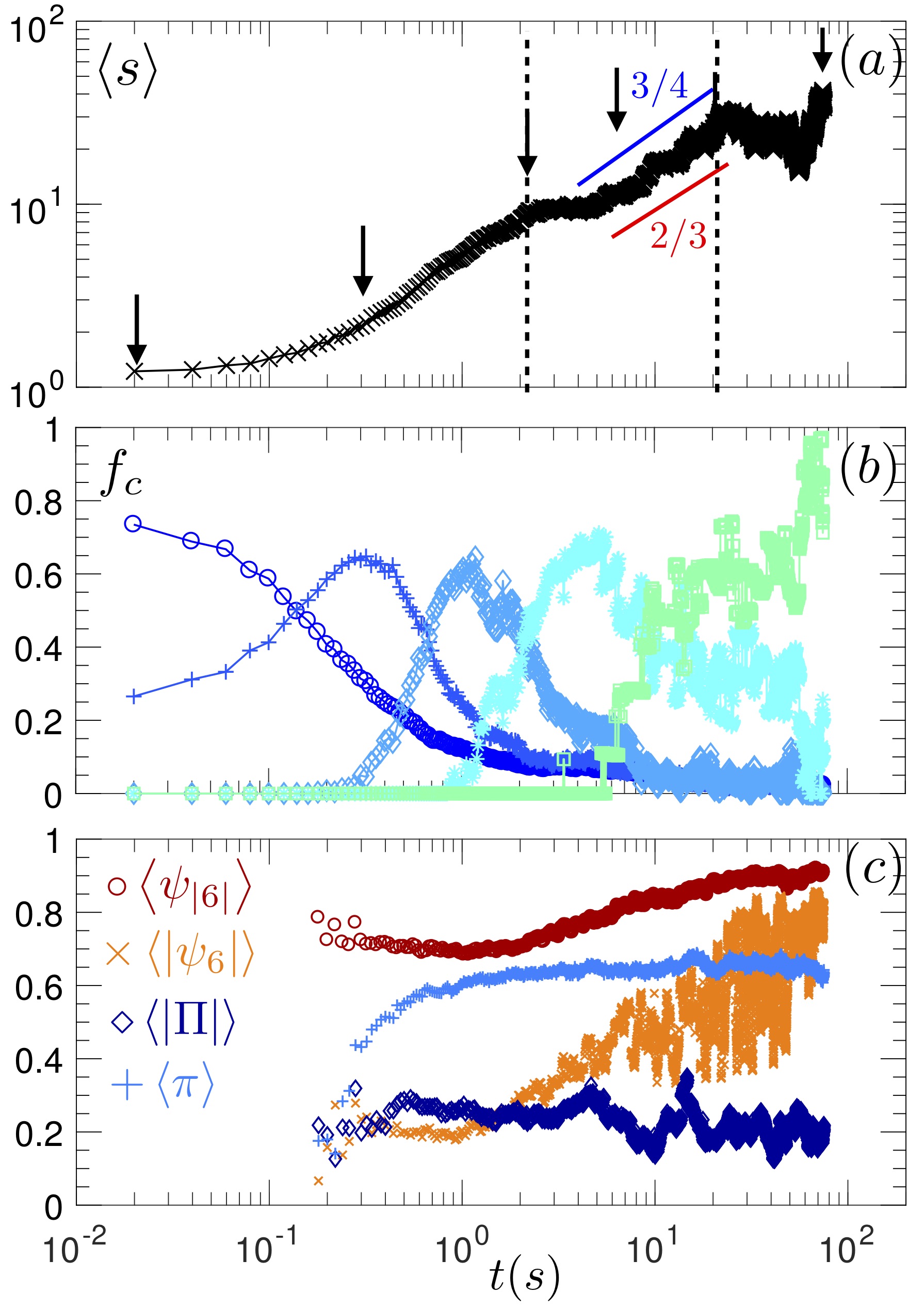}
\vspace{-0mm}
\caption{{\bf Cluster size and order parameters:} (a) Average cluster size versus time; the arrows point at the times of the snapshots shown on Fig.~\ref{fig:coarsening}; the vertical dashed lines separate the three dynamical regimes described in this Letter. (b) Fraction of particles inside clusters of size $(\circ)\,s = 1, (+)\, s\in[2, 19], (\diamond)\, s\in[20, 99], (*)\, s\in[100, 499], (\square)\, s\in[500, 4999]$. (c) Weighted average of the hexagonal $\psi_{|6|}$, hexatic $\left|\psi_6\right|$, aligning $\pi$, and polar $\left|\PPi\right|$, order parameters (see text for definitions).}
\label{fig:growth}
\vspace{-5mm}
\end{figure}

%As illustrated in Fig.~\ref{fig:coarsening}, particle aggregation starts right at the onset of self-propulsion. 
Fig.~\ref{fig:growth}-(a,b) display the average cluster size and the fraction of particles inside clusters of increasing size. The average cluster size is
$\left<s\right>=\frac{1}{N(t)}\sum_i s_i$, with $N(t)$ the number of clusters at time $t$ and $s_i$ the number of colloids inside cluster $i$.  
One readily distinguishes three regimes. At short time $t<2s$, the dynamics are dominated by the aggregation of isolated particles  in small clusters of average size $\left<s\right> < 10$ and maximal size $s_{max}\simeq 250$ (regime I). This first regime ends when most of the individual particles have aggregated. It is marked by an abrupt slowing down of the aggregation: $\left<s(t)\right>$ remains flat for another $2$s, before aggregation restarts in the form of the coarsening of the previously formed clusters (regime II). This coarsening process is itself interrupted at longer times ($t=20$s) leading to a final regime dominated by strong fluctuations of the average cluster size around $\left<s\right> = 30$, with $s_{max}\simeq 2000$ (regime III). 

% -- FIGURE 0002 ------------------------------
When the clusters form, they rapidly develop hexagonal order, and polar alignment of the particles.  Fig~\ref{fig:growth}-(c) reveals how structural and polar order develops. For each cluster of size $s$, the structural order is characterized using the hexagonal, respectively hexatic, order parameter $\psi_{|6|}=\frac{1}{s}\sum_{k=1}^{s} \left| \psi_{6,k}\right|$, resp. $\left|\psi_6\right|=\left|\frac{1}{s}\sum_{k=1}^{s} \psi_{6,k}\right|$, where $\psi_{6,k} = \frac{1}{N_k}\sum_{j=1}^{N_k} \exp (6i\theta_{jk})$, with $\theta_{jk}$ the orientation of the link connecting two neighboring particles, and the sum runs over the $N_k$ nearest neighbors of particle $k$, using a cut-off distance of $1.2d$.
The alignment is defined at the particle level as $\pi_k = \frac{1}{N_k}\sum_{j=1}^{N_k} \nn_j\!\cdot\!\nn_k$ and at the cluster level as $\pi = \frac{1}{s}\sum_{k=1}^{s} \pi_k$.
The polarity of a cluster of size $s$ is given by $\left|\PPi\right|=\left |\frac{1}{s}\sum_{k=1}^{s} \nn_k \right|$.
Fig~\ref{fig:growth}-(c) is obtained by averaging over clusters with $s\geq7$ present at time $t$, weighting the average with the cluster size. One readily sees that the different orders develop at different pace.
In the following we shall describe the three growth regimes, focusing on the interplay between structure, polar ordering and growth. In the first regime, we base our analysis on the statistics of the cluster size. At longer times we concentrate on the structure and orientational organization of the clusters to identify the reasons for the interruption of the coarsening process.
%
% -- FIGURE 0003 ------------------------------
\begin{figure}[t] 
\center
\vspace{0cm}
\includegraphics[width=\columnwidth]{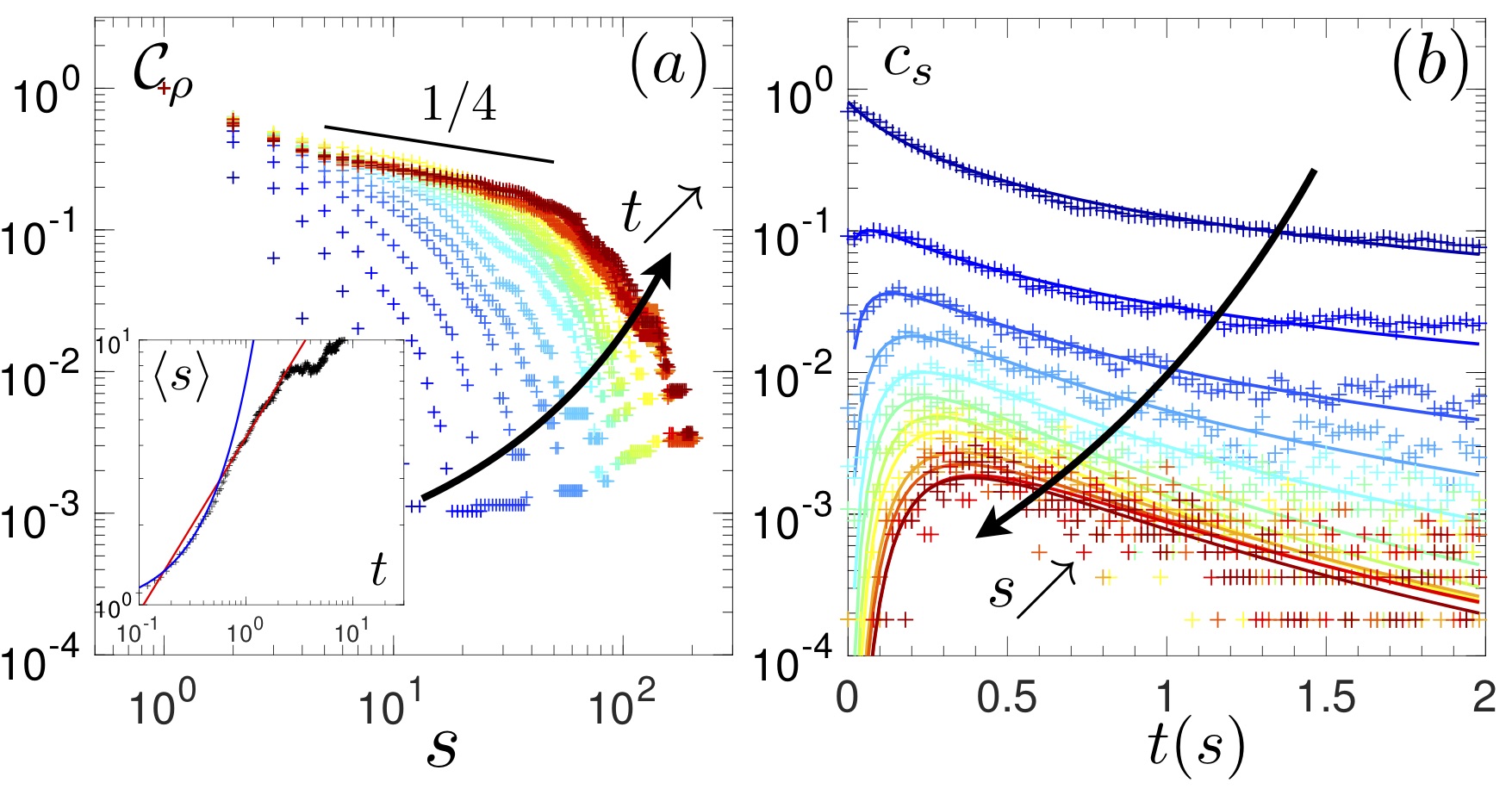}
\vspace{-3mm}
\caption{{\bf Short-time aggregation:} (a) Cumulative distribution of the cluster size, $\mathcal{C}_{\rho}(s,t)$, for increasing times $t\in[0, 2]$~s every $0.2$~s (from blue to red); inset : $\left<s(t)\right>$; the continuous blue line is an exponential fit $e^{t/t_1}$, with $t_1=25s$; the continuous red line is a power law fit $t^\gamma$, with $\gamma = 2/3$. (b) Cluster size histogram $c_s(t)$ as a function of time, for $s\in[1, 12]$; the continuous lines are fits of the form $h(t) \propto \frac{(t/t_2)^{p(s)}}{(1+t/t2)^{q(s)}}$, with $t_2=10s$, $p(s)= 0.6\,(s-1)$ and $q(s)=s-1$  (see text for details).}
\label{fig:shorttime}
\vspace{-0.5cm}
\end{figure}

{\it --- Short-time dynamics ---} 
The initial aggregation follows a standard route, akin to equilibrium aggregation: the cluster size distribution $\rho(s,t)$ is exponential at very short times and progressively develops a power law regime. This is best illustrated by the cumulative distribution $\mathcal{C}_{\rho}(s,t)=\int_s^{\infty}du\rho(u,t)$ plotted at successive times on Fig.~\ref{fig:shorttime}-(a), from which we infer that: ${\rho}(s,t) \propto s^{-\alpha} \exp(-s/s^*(t))$, with $\alpha \simeq 1.25 $, smaller than the typical values $\alpha\in~[1.7, 2]$~\cite{Peruani:2012dya,Levis:2014ina,Redner:2016cc,Ginot:2018ff}, indicating a truly broad distribution of sizes.
The crossover size $s^*$ sets the average cluster size $\left<s\right>$, the evolution of which is displayed in the inset. The initial exponential growth, expected for an aggregation instability, coincides with the formation of branched clusters (see Fig~(\ref{fig:coarsening})-b), suggestive of a diffusion limited process (DLA). This initial regime is followed by a power law growth of the cluster size $\left<s\right> \sim t^{\gamma_1}$, with $\gamma_1 \simeq 2/3$, during which the clusters rapidly become rather compact. The characteristic length $\mathcal{L}$ associated with the clusters growth thus follows $\mathcal{L}\sim t^{1/3}$, as prescribed by the Cahn-Hilliard equation, which describes the simplest from of phase separation for a conserved field~\cite{Cahn:1958ds}. 
The short-time dynamics can be further characterized by the cluster size histogram $c_s(t)=n_s(t)/M$, with $n_s(t)$ the number of clusters of size $s$, displayed in Fig~(\ref{fig:shorttime}-b). Assuming constant rate aggregation among clusters, one can show that, starting with an initial state only composed of individual particles, $c_s(t)=\frac{(t/t_2)^{p(s)}}{(1+t/t_2)^{q(s)}}$, with $p(s)=s-1$ and $q(s)=s+1$~\cite{Krapivsky:1311513}. Here we find $p(s) \simeq 0.6 (s-1)$ and $q(s) \simeq s-1$. The most interesting difference is that for $t\gg 1$, $c_s(t)$ decreases much slower than the prescribed $1/t^2$, indicating that the aggregation process starts competing with evaporation and/or fragmentation events.

\begin{figure}[t] 
\center
\vspace{0cm}
\includegraphics[width=0.95\columnwidth]{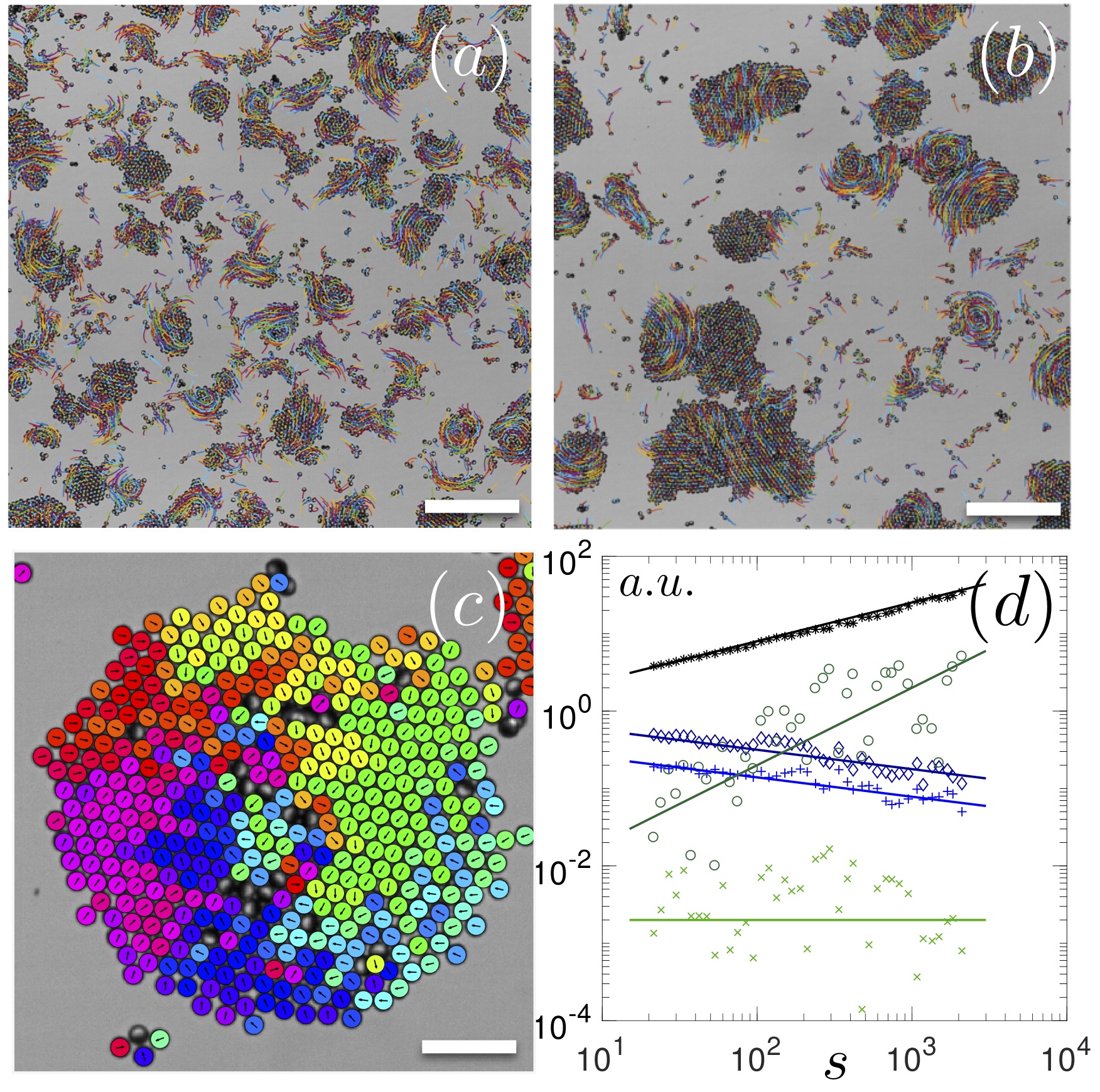}
\vspace{-0mm}
\caption{{\bf Coarsening of active clusters:} (a)-(b) Snapshots at times $t = 8$s and $t = 42$s, with traces of the particles integrated over $\Delta t = 0.4$s. Scale bar is $100\mu$m. (c) Zoom on the colloids orientations inside one cluster. Scale bar is $25\mu$m.(d) Cluster properties (in arbitrary units) as a function of cluster size $s$: $(*)$ radius of gyration $R_G$, $(+)$ speed $\left|\bar{v}\right|$, $(\diamond)$ polarity $\left| \PPi \right|$, $(\times)$ angular velocity $\omega$, $(\circ)$ torque lever $\tau/f_0$.}
\label{fig:activeclusters}
\vspace{-5mm}
\end{figure}

{\it --- Aggregation ---}
The onset of the second dynamical regime is initially marked by the slowing down of the aggregation process (Fig~\ref{fig:growth}-a): for $t>2$~s, most of the particles are already trapped within clusters but point in random directions (Fig~\ref{fig:coarsening}-c). This however does not last for long: after another $2$~s, new dynamics set in (   Fig~(\ref{fig:activeclusters}-(a,b)): particles within clusters have locally aligned their orientation with their neighbors (see Figs.~\ref{fig:growth}-(c) and~\ref{fig:activeclusters}-(c)), building up spatial correlations at the scale of the cluster size. As a result, the clusters are animated with quasi-rigid-body translational and rotational motion. The dynamics are heterogeneous -- some clusters are static, other translate almost at the nominal speed of the individual colloids and others spin, like rigid bodies -- and highly intermittent because collisions among the clusters redistribute the alignment of the colloids.

% -- FIGURE 0004 ------------------------------
For a given cluster of size $s$ at time $t$, we measure the velocity of each colloid $\vv_k = \left(\rr_k(t+\Delta t) - \rr_k(t)\right)/\Delta t$, with $\Delta t = 0.02$~s, and subsequently extract the position $\bar{\rr} = \frac{1}{s}\sum_{k=1}^{s} \rr_k$, velocity $\bar{\vv} = \frac{1}{s}\sum_{k=1}^{s} \vv_k$ of the center of mass, the radius of gyration $R_G=\left(\frac{1}{s}\sum_{k=1}^{s} \left|\rr_k-\bar{\rr}\right|^2 \right)^{1/2}$ and the
absolute angular rotation $\omega = \left| \frac{1}{s}\sum_{k=1}^{s} \frac{(\rr_k-\bar{\rr}) \times \vv_k}{\left|\rr_k-\bar{\rr}\right|^2} \right|$.
Here we assume each colloid exerts a force $f_0 \nn_k$ on the cluster it belongs to. Then the amplitude of the mean force exerted on a cluster of size $s$ is simply $f = f_0 \left| \PPi \right|$ and the amplitude of the mean torque is $\tau = f_0\left|  \frac{1}{s}\sum_{k=1}^{s} (\rr_k-\bar{\rr}) \times \nn_k \right|$.
Figure~\ref{fig:activeclusters}-(d), shows how the radius of gyration, the translational and angular velocity, the mean force and the mean torque scale with the cluster size: $R_G\sim s^{0.5}$, $v\sim s^{-0.25}$, $\omega\sim s^{0.0}$, $f\sim s^{-0.25}$ and $\tau\sim s^{1.0}$. As a result, the translational drag coefficient $\xi_t = \frac{s f}{v} \sim s^{1.0}$ while the rotational drag  $\xi_r = \frac{s \tau}{\omega} \sim s^{2.0}$. Both scalings contrast with the Stokes prediction for a disk, ($\xi_t\sim s^0 + \log \text{corrections}$ and $\xi_r\sim s$), prohibiting the description of the cluster as a simple solid disk. The obtained scalings are however in agreement with the cumulative drag model proposed to describe active clusters~\cite{Ginot:2018ff}.
These scaling laws are of crucial importance since they set the collision frequency amongst clusters, and thereby the temporal scaling of the coarsening dynamics. In the present case, the average cluster size $\left<s\right> \sim t^{\gamma_2}$, with $\gamma_2\in[2/3, 3/4]$ (see. Fig~(\ref{fig:growth})-a). 

On the theoretical side, proceeding further would require solving the master equation governing the probability density of cluster sizes. There are very few cases where it can be solved exactly and one often restricts the description to the '‘monomer approximation’', at least for the fragmentation process~\cite{Peruani:2013eca,Ginot:2018ff}. It is however clear that the present dynamics, which mainly involve cluster-cluster processes, would not be captured within such an approximation. Furthermore, the long time scales dynamics never take place, as we shall now see that the phase separation is anyway interrupted.

\begin{figure}[t] 
\center
\vspace{0cm}
\includegraphics[width=0.95\columnwidth]{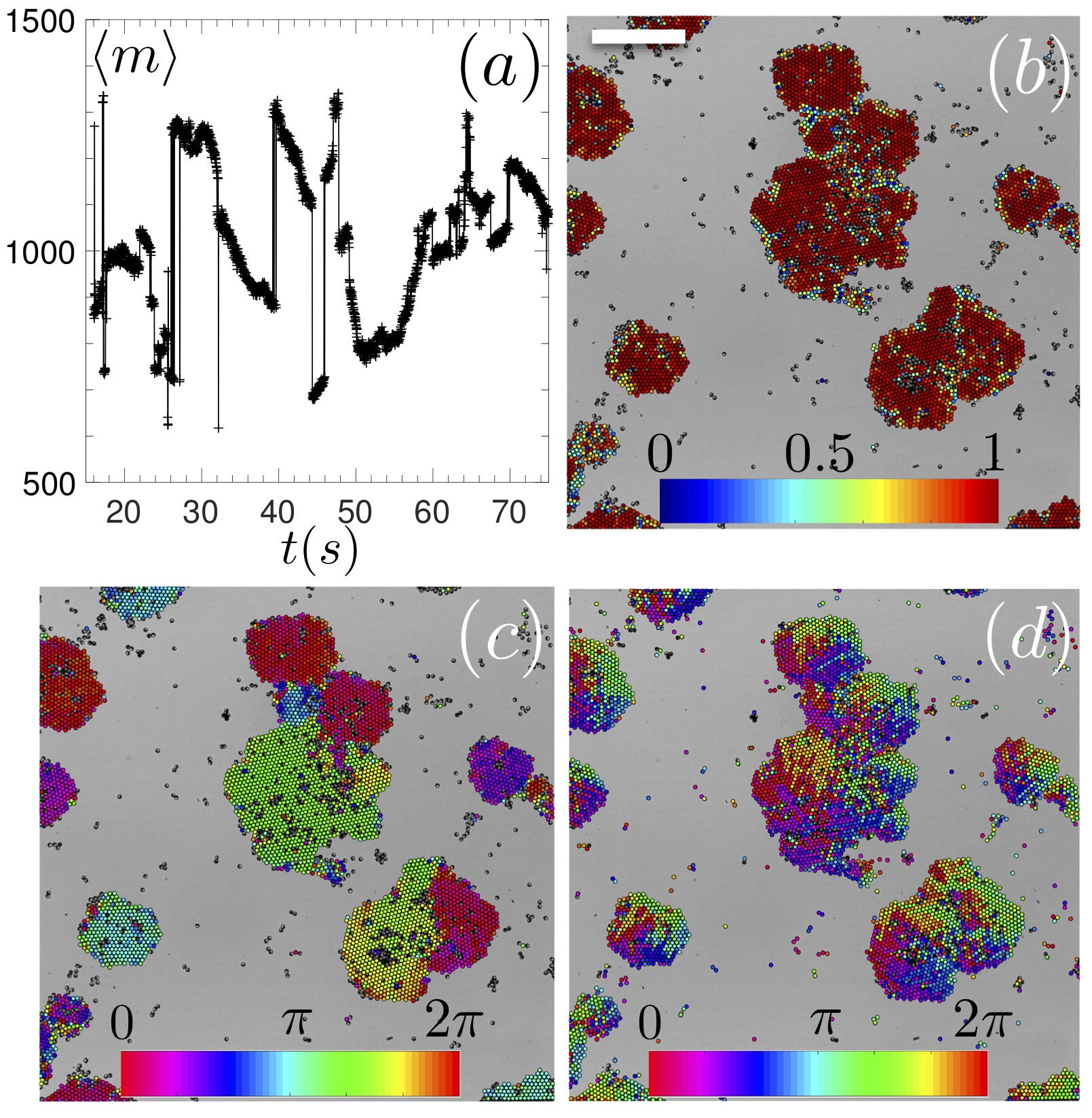}
\vspace{-0mm}
\caption{{\bf Ordering within clusters:} (a) Temporal evolution of the mass-weighted average of the cluster size.  
Large clusters observed at time $t=20$s with particles color-coded by (b) the amplitude, (c) the argument of the complex local bond-orientational order parameter $\psi_6$ and (d) the orientation of their polar axis. Scale bar is $100\mu$m. See also Movie-2,3,4 in Supp. Mat.}
\label{fig:clusters}
\vspace{-0.5cm}
\end{figure}

{\it --- Interrupted Phase Separation ---}
At long times, one would expect that most colloids aggregate into a few very large clusters ($s\simeq 1000$), which eventually merge and form one dense droplet surrounded by a very dilute gas of individual colloids. Coarsening would then saturate because of the finite number $M$ of colloids. The dynamics are actually far more complex, as evidenced by the large fluctuations observed in the temporal evolution of the mass-weighted average of the cluster size $\left<m\right>(t) = \frac{\sum_i s_i^2}{\sum_i s_i}$  (Fig.~\ref{fig:clusters}-(a)). Frequent very sharp breaking events take place, which, as we shall now argue, result from the imperfect aggregation of the clusters beyond a certain size.

The largest clusters exhibit a significant hexagonal order (large values of $\left|\psi_{6,k}\right|$), as expected for dense assemblies of spherical particles in 2D (Fig.~\ref{fig:clusters}-(b)). However, one can identify sub-regions with high hexagonal order separated by domain walls with low structural order. This is further confirmed by considering the orientation of the local hexagonal lattice, as provided by the argument of $\psi_{6,k}$, shown in Fig.~\ref{fig:clusters}-(c): domains with different orientations of the hexagonal order coexist within a cluster. These domains are inherited from the aggregation history of the cluster: each orientational domain corresponds to a recently aggregated cluster. 
%
% -- FIGURE 0005 ------------------------------
The structural defects composing the domain walls will migrate towards the cluster boundaries, until a homogeneous structural state is eventually reached. This is however a slow process as evidenced by the slow growth of the hexatic order parameter $\left<\left|\psi_6\right|\right>$ (Fig.~\ref{fig:growth}-(c)). It is interrupted by the intense shear induced by the rotational and, to a lesser extent, translational motions of the clusters described in the previous section. Fig.~\ref{fig:clusters}-(d) verifies that the structurally homogeneous domains indeed show the spatial correlations in particle alignment that are responsible for such motions. Note that the shear induced by the rotation of two adjacent clusters is typically localized to one colloidal layer. Hence the shear rate increases with the cluster size like $ R_G \omega$. On the contrary, we expect that the rate at which defects leave the grain boundary, and thus at which the large clusters heal, decrease with the cluster size. A critical size above which coarsening is interrupted is then always reached, irrespective of the details of the healing mechanism.

In summary, the coarsening dynamics result from the competition of three types of dynamics, that of Motility Induced Phase Separation, that of structural ordering and that of polar alignment of the particles. The MIPS dynamics is initially the fastest one and rapidly leads to the formation of dense and compact clusters during the first regime. As long as the clusters are not too large, the structural ordering dynamics is fast enough compared to the aggregation rate and the newly formed clusters rapidly become structurally homogeneous. Simultaneously, slower polar ordering develops spatial correlations at the cluster scale. These correlations are responsible for the presence of torques and forces, which in turn ensure the motility of the clusters and thereby set the aggregation rate during the second regime. In the last regime, the clusters have reached sizes such that the structural ordering now competes with the stresses inherited from the misalignment within the aggregating clusters. In this situation the long time state is very much reminiscent of the traffic jam and gliders reported in a simulations of active rods~\cite{Peruani:2011kwa}. Depending on the relative growth rate of the three types of dynamics, one may expect different asymptotic states, leaving space for yet unexplored collective organizations.  

{\it --- Acknowledgments ---}
We thank Chantal Valeriani and Julien Tailleur for inspiring discussions and Jeroen Rijks for his contribution to the initial research. MNvdL acknowledges support from H2020 Marie Sklodowska-Curie Individual Fellowship "TOPACT".
\vspace{-2mm}
\bibliography{Active.bib}

% \begin{thebibliography}{0}
% \end{thebibliography}

\end{document}